\documentstyle[12pt,fleqn]{article}
\def\beeq{\begin{equation}}
\def\eneq{\end{equation}}
\def\beeqa{\begin{eqnarray}}
\def\eneqa{\end{eqnarray}}

\addtocounter{section}{-1}
\begin{document}

\begin{center}

\vspace{2cm}

{\large {\bf {
Mechanism of magnetism in stacked nanographite\\
with open shell electrons
} } }

\vspace{1cm}

{\rm Kikuo Harigaya$^{1,2,}$\footnote[1]{E-mail address:
\verb+k.harigaya@aist.go.jp+;
URL: \verb+http://staff.aist.go.jp/k.harigaya/+}}
and Toshiaki Enoki$^3$

\vspace{1cm}

$^1${\sl National Institute of Advanced Industrial
Science and Technology (AIST),\\
Umezono 1-1-1, Tsukuba 305-8568, Japan}
\footnote[2]{Corresponding address}\\
$^2${\sl Interactive Research Center of Science,
Tokyo Institute of Technology,\\
Oh-okayama 2-12-1, Meguro-ku, Tokyo 152-8551, Japan}\\
$^3${\sl Department of Chemistry,
Tokyo Institute of Technology,\\
Oh-okayama 2-12-1, Meguro-ku, Tokyo 152-8551, Japan}

\end{center}

\vspace{1cm}

\noindent
{\bf Abstract}\\
Antiferromagnetism in stacked nanographite is investigated 
with using the Hubbard-type models.  The A-B stacking 
or the stacking near to that of A-B type is favorable
for the hexagonal nanographite with zigzag edges, in order 
that magnetism appears.  Next, we find that the open shell 
electronic structure can be an origin of the decreasing 
magnetic moment with the decrease of the inter-graphene
distance, as experiments on adsorption of molecules suggest.

\mbox{}

\noindent
PACS numbers: 75.30.-m, 75.70.Cn, 75.10.Lp, 75.40.Mg

\mbox{}

\pagebreak

\section{Introduction}

Nanographite systems are composed of the structural units with 
stacking of graphene sheets of the nanometer size [1].
They show novel magnetic properties, such as, spin-glass 
like behaviors [2], and the change of ESR line widths 
while gas adsorptions [3]. Recently, it has been found [4,5] 
that magnetic moments decrease with the decrease of the 
interlayer distance while water molecules are attached 
physically.

In the previous papers [6,7], we have considered the 
stacking effects of hexagonal nanographite layers
in order to investigate mechanisms of 
antiferromagnetism using the Hubbard-type model with 
the interlayer hopping integrals and the onsite 
repulsion $U$.  We have taken account of the 
nanographite systems where the numbers of the sites
and electrons in a layer are same, and therefore the 
electronic systems of the layer have closed shell structures.  
In the calculations, the finite magnetization develops,
as the hopping interactions between layers increase 
in the case of the A-B stacking [6].  The same conclusion
has been obtained when the system is near the A-B stacking
and the interlayer distance becomes shorter [7].
The A-B stacking should exist in nanographite systems,
because the exotic magnetisms have been observed
in recent experiments [1-3].  The decrease of the 
interlayer distance while attachment of water molecules 
makes $t_1$ larger.  However, it is known that the 
magnetism decreases while the attachment of molecules [4,5].
The calculation for the closed electron systems
cannot explain the experiments even qualitatively.

The purpose of this paper is to extend the previous
calculations to the systems with open shell electronic
structures in the isolated nanographite layer.
There are several candidates for the open shell
of electronic structures.  One of them is the
electron number changes due to the presence of the
active side groups.  The electron or hole is donated
from the side group.  The effects can be modeled
by a site potential in the idea of the model hamiltonians.
This idea will be investigated in the present paper.
The other origin of the open shell is a geometrical
origin.  The phenalenyl molecule C$_{13}$ with three 
hexagonal rings [8,9], and the triangulene C$_{22}$ 
with six hexagons [10,11] are the examples of 
small graphene layers which have open shell electronic
structures.  The stacking effects of such kinds
of molecules will be reported elsewhere [12].

We shall study with the two kinds of models which have
been used in the papers [6,7].  Site potentials which 
simulate the additional side groups are introduced in the 
Hubbard-type hamiltonians.  The change of the electron number
is taken into account, too.  We will show that
the A-B stacking or the relative interlayer 
relation near the A-B stacking is favorable 
for the hexagonal nanographite with zigzag edges, 
in order that magnetism appears.  We also find 
that the open shell electronic structure can be 
an origin of the decreasing magnetic moment 
with adsorption of molecules.

In the next section, we explain our models and
review the method of the numerical calculations.
In sections 3 and 4, the results of the two
models are reported and discussed.  The paper
is closed with a summary in section 5.

\section{Models}

First, we will investigate with the {\bf model I} [6], 
\beeqa
H &=& -t \sum_{\langle i,j \rangle: {\rm intralayer}} \sum_\sigma
(c_{i,\sigma}^\dagger c_{j,\sigma} + {\rm h.c.}) \nonumber \\
&-& t_1 \sum_{\langle i,j \rangle: {\rm interlayer}} \sum_\sigma
(c_{i,\sigma}^\dagger c_{j,\sigma} + {\rm h.c.}) \nonumber \\
&+& U \sum_i n_{i,\uparrow} n_{i,\downarrow},
\eneqa
where $n_{i,\sigma} = c_{i,\sigma}^\dagger c_{i,\sigma}$ for 
$\sigma = \uparrow$ and $\downarrow$; $c_{i,\sigma}$ is
an annihilation operator of an electron at the $i$th site
with spin $\sigma$; the sum of the first line is taken
over the nearest neighbor pairs $\langle i,j \rangle$
in a single layer of the nanographite; the sum of the
second line is taken over sites where the distance 
between two positions of the neighboring layers is
shortest; $t_1$ is the strength of the weak hopping 
interaction between neighboring layers; the positions 
of $t_1$ are shown by the filled circles in Fig. 1 (a); 
and the last term of the hamiltonian is the 
strong onsite repulsion with the strength $U$.  We 
will vary the strength in a reasonable range.  The actual 
values of $U$ for carbon atoms are of the similar 
order of magnitudes as in this paper: for example, we have 
found $U=4t$ for the neutral C$_{60}$ and C$_{70}$ [13]
and $U=2t$ for the doped C$_{60}$ and C$_{70}$ [14]
in the theoretical characterizations of the
optical absorption experiments.

Second, we will study with the {\bf model II} [7],
\beeqa
H &=& -t \sum_{\langle i,j \rangle: {\rm intralayer}} \sum_\sigma
(c_{i,\sigma}^\dagger c_{j,\sigma} + {\rm h.c.}) \nonumber \\
&-& \sum_{(i,j): {\rm interlayer}} \sum_\sigma
\beta (r_{i,j})
(c_{i,\sigma}^\dagger c_{j,\sigma} + {\rm h.c.}) \nonumber \\
&+& U \sum_i n_{i,\uparrow} n_{i,\downarrow},
\eneqa
where the sum of the
second line is taken over pairs of the sites $(i,j)$
in neighboring layers with a cutoff for long distance as used 
in [15] of the calculation for multiwall carbon nanotubes; 
the function $\beta (r)$ is given by 
\beeq
\beta(r) = A {\rm exp} (-r/\zeta)
\eneq
where $r$ is the distance between carbon atoms, $A = 5.21t$,
and $\zeta = 0.86$\AA.  The magnitude, $\beta(r=3.40$\AA$)
= 0.1t$, is a typical value for the interlayer interaction
strength in the tight binding model for A-B stacked graphite 
layers [16]:  the explicit value is about 0.35 - 0.39 eV, 
and $t \sim 3$eV gives the interaction strength about $0.1t$.

We investigate the systems which have open shell electronic 
structures when a nanographene layer is isolated.  The 
effects of additional charges coming from functional side 
groups are simulated with introducing site potentials 
$E_s$ [17,18] at edge sites in the model I or II.  
The form of the site potentials is
\beeq
H_{\rm site} = E_s \sum_{i \in I}
\sum_\sigma c_{i,\sigma}^\dagger
c_{i,\sigma},
\eneq
where the sum of $i$ is taken over the set of the positions 
of site potentials $I$.  When $E_s > 0$, the electron number
decreases from the average value at the site $i$, and 
the site potential means the electron attractive groups.
When $E_s < 0$, the electron donative groups are simulated
because of the increase of the electron number at the
site potentials.  Here, we take $E_s = -2t$, and one
additional electron per layer is taken account.

The finite size system, whose number of the stacking 
layers is similar to that of the samples with the 
nanometer size, is solved numerically with using the 
periodic boundary condition for the stacking direction, 
and we obtain two kinds of solutions.  One of them 
is an antiferromagnetic solution, where the number 
of up spin electrons is larger than that of down spin 
electrons in the first layer, the number of down spin 
electrons is larger than that of the up spin electrons 
in the second layer, and so on.  The other kind of solution
is a nonmagnetic solution.  The present author has 
discussed the antiferromagnetism in C$_{60}$ polymers, 
too [19].  The same technique (unrestricted Hartree-Fock 
approximation) used in ref. [6,7,19] is effective for 
the open shell electronic systems of this paper.
The number of unit cells in the stacking direction is 10,
and therefore there are 20 layers in the system
used for the numerical treatment.

In the {\bf model I}, the parameters are changed 
within $0 \leq t_1 \leq 0.5t$ and $0 \leq U \leq 4t$.  
The realistic value of $t_1$ is estimated to be 
about $0.1t$ at most [16], but we change this 
parameter for more extended regions in order to 
look at the behaviors of solutions in detail.  In the 
antiferromagnetic solutions, the number of electrons is 
the same with the number of sites, and electronic states
are half-filled.  Because we assume the electron
donative case $E_s < 0$, the number of electrons per
layer is larger than the site number.  Here, we take
the electron number per layer $n_{\rm el} = 25$
and the site number per layer $n_{\rm site} = 24$.
All of the quantities of the energy dimension are 
reported using the unit $t$ ($\sim 2.0 - 3.0$ eV).

In the {\bf model II}, we investigate the continuous 
change between the A-B stacking [Fig. 1 (a)] and
the A-A stacking [Fig. 1 (b)].  We will move the
first layer of Fig. 1 (b) to the upper direction.  
When the relative shift $d=0$, the geometry is of 
Fig. 1 (b).  As $d$ becomes larger, the system 
changes from the A-A stacking to the A-B stacking.  
When $d = a$ ($a$ is the bond length in each layer), 
the system has the geometry of Fig. 1 (a).  In 
increasing $d$, the edge sites feel weaker 
interactions from neighboring layers.  Such the
changes will give rise to variations in magnetic properties.
The magnetism of the model II will be discussed with 
varying $d$ and $R$ ($R$ is the interlayer distance)
in section 4.

\section{Results of model I}

Here, we consider the Hubbard-type model for systems which 
have open shell electronic structures when a nanographene 
layer is isolated.  The effects of additional charges 
coming from functional side groups are considered
using the {\bf model I} whose form is given by eq. (1).

Figure 2 displays the absolute values of total magnetic
moment per layer.  Figures 2 (a), (b), and (c) are for
the A-B stacking, and Fig. 2 (d) represents the case
of the A-A stacking.  The strengths of the Coulomb 
interaction $U$ are shown in the figure captions.
In Fig. 2 (a), the site potentials 
locate at the site D in the first layer [Fig. 1 (a)], 
and at the symmetrically equivalent site D' in the second layer.  
The site potentials exist at the sites E and E' in 
Fig. 2 (b), and they are present at the sites F and F' 
in Fig. 2 (c).  The total magnetization is a decreasing 
function in these three figures.  The decrease is 
faster in Figs. 2 (b) and (c) than in Fig. 2 (a).  
The sites E and F are neighboring to the site with 
the interaction $t_1$, and thus the localized
character of the magnetic moment can be affected easily
in these cases.  The decease of magnetization by the
magnitude $30-40$\% with the water molecule attachment [3]
may correspond to the case of Fig. 2 (b) or Fig. 2 (c).

Next, we note that the magnetic ordering is not 
present for the A-A stacking case [shown in Fig. 1 (b)]
whose numerical result is shown in Fig. 2 (d).  
This is owing to the fact that all the sites between
the neighboring layers interact via the hopping integrals, 
and that the itinerant characters of the electrons are 
dominant strongly for the A-A stacking.  The present 
results are in agreement with the calculations for the 
closed shell systems reported in the previous paper [6].

\section{Results of model II}

In this section, the effects of site potentials and 
additional charges are investigated with the {\bf model II} 
taking into account of the continuous relative shift 
between layers.  The model eq. (2) has been used in [7], 
and is reviewed in the section 2 of this paper.

Figure 3 shows the absolute magnitude of the total
magnetic moment per layer as a function of the relative
shift $d$.  The site potentials with the strength 
$E_s = - 2t$ are present at the sites D (E, and F)
in the first layer, and at the sites D' (E', and F')
in the second layer, in Figs. 3 (a) [(b), and (c)],
respectively.  Here, the interlayer distance $R$ is 
fixed and the Coulomb strength $U$ is varied in the 
series of the plots.  The system is with the A-A stacking
at $d=0$, and the stacking is of the A-B type at
$d=a$.  At $d=0$ and in the smaller region of $d$,
there is not finite magnetization due to the strong
interlayer hopping interactions near the A-A stackings.
This qualitative property agrees with that of the
model I and also with the calculation of the closed
shell electron system [7].  However, there appear
the finite magnetic moments with the antiferromagnetic 
sign alternation in the stacking direction for the
region $d > 0.5a$ and for $U$ larger than a critical
value.  The onset of the magnetic moment at $d=0.5a$
might be due to the presence of a cutoff of the
long distance interaction, which has been used 
in [15], also.  The magnetic moment increases as
$U$ becomes larger as expected for the increase
of the localized characters of electrons.  In contrast,
the magnetic moment is a weak decreasing function with
respect to $d$.

Figure 4 shows the same quantity as a function of $d$.  
In these plots, the Coulomb interactions are taken constant 
as a representative strength $U = 1.2t$, and the interlayer
distance $R$ is varied.  The interlayer distance
becomes smaller from the filled squares, through
the open squares, to the filled circles.  In the 
regions of $d$ where the finite magnetization is
present, the magnetic moment decreases strongly
while the interlayer distance becomes shorter.
The total hopping interactions become larger
as $R$ decreases.  This effect enhances the itinerant
characters of electrons in the stacking direction.
Therefore, the magnetic moment becomes smaller with 
respect to the decrease of the interlayer distance.
Such the qualitative behavior of the open shell
electron system is in contrast with that of the
closed shell electron system reported in the paper [7].
The present calculation is in qualitative 
agreement with that of the experiment which reports 
the decrease of the magnetization in the course
of the adsorption of molecules [4,5].  In fact,
the decrease of the interlayer distance about
0.4 \AA~ while the adsorption of water molecules
has been reported [20].

The magnetic moment decreases easily for the site
potentials at E and E' [Figs. 3 (b) and 4 (b)],
and also for the site potentials at F and F's
[Figs. 3 (c) and 4 (c)].  It does not vary so
much in the case of the site potentials at D and
D' [Figs. 3 (a) and 4 (a)].  Such the quantitative
difference comes from the contrast whether the site
potentials are far from the sites with strong
interlayer hopping interactions or not.  The difference
of the effects of the site potentials at D (D') from 
those of the potentials at E (E') or F (F') agrees 
qualitatively with the result of the model I reported 
in the previous section.

To conclude, the two calculations of models I and II 
of this paper agree with the experiments, qualitatively.  
The case of the site potential at E or F agrees with 
experiments, even quantitatively.  We can explain 
the decrease of magnetism of $30-40$ \% 
in the process of adsorption of molecules [4,5].
Therefore, the open shell electronic structure due to 
the active side groups is a good candidate which 
could explain the exotic magnetisms.  The exotic 
magnetism is related with the edge states which
have large amplitudes at the zigzag edge atoms
of the each graphene layer.  The important roles 
of the edge states have been discussed intensively
for the nanographite ribbon systems recently [21-23].

\section{Summary}

Antiferromagnetism in stacked nanographite has been 
investigated with the Hubbard-type models.  The A-B 
stacking or the relative interlayer relation near
the A-B stacking is favorable for the hexagonal 
nanographite with zigzag edges, in order that magnetism 
appears.  We have also found that the open shell electronic 
structure can be an origin of the decreasing magnetic 
moment with adsorption of molecules.

\mbox{}

\begin{flushleft}
{\bf Acknowledgements}
\end{flushleft}

\noindent
The authors are grateful for interesting discussion with
N. Kawatsu, H. Sato, K. Takai, T. Ohshima, Y. Miyamoto, 
K. Kusakabe, K. Nakada, K. Wakabayashi, and M. Igami.  
Useful discussion with the members of Nanomaterials
Theory Group, Nanotechnology Research Institute, AIST
is acknowledged, too.

\pagebreak
\begin{flushleft}
{\bf References}
\end{flushleft}

\noindent
$[1]$ M. S. Dresselhaus, "Supercarbon: Synthesis, Properties
and Applications", eds. S. Yoshimura and R. P. H. Chang,
(Springer Verlag, Berlin, 1998), Part II.\\
$[2]$ Y. Shibayama, H. Sato, T. Enoki, and M. Endo, 
Phys. Rev. Lett. {\bf 84}, 1744 (2000).\\
$[3]$ N. Kobayashi, T. Enoki, C. Ishii, K. Kaneko, and M. Endo,
J. Chem. Phys. {\bf 109}, 1983 (1998).\\
$[4]$ N. Kawatsu, H. Sato, T. Enoki, M. Endo, 
R. Kobori, S. Maruyama, and K. Kaneko,
Meeting Abstracts of the Physical Society of Japan
{\bf 55} Issue 1, 717 (2000).\\
$[5]$ N. Kawatsu, H. Sato, T. Enoki, M. Endo, 
R. Kobori, S. Maruyama, and K. Kaneko, preprint.\\
$[6]$ K. Harigaya, J. Phys: Condens. Matter {\bf 13}, 
1295 (2001).\\
$[7]$ K. Harigaya, Chem. Phys. Lett. {\bf 340}, 123 (2001).\\
$[8]$ K. Fukui et al, Synth. Metals {\bf 103},
2257 (1999).\\
$[9]$ K. Fukui et al, Mol. Cryst. Liq. Cryst.
{\bf 334}, 49 (1999).\\
$[10]$ G. Allinson, R. J. Bushby, and J. L. Paillaud,
J. Am. Chem. Soc. {\bf 115}, 2062 (1993).\\
$[11]$ M. J. Bearpark, M. A. Robb, F. Bernardi, and
M. Olivucci, Chem. Phys. Lett. {\bf 217}, 513 (1994).\\
$[12]$ K. Harigaya, N. Kawatsu, and T. Enoki,
"Nanonetwork Materials: Fullerenes, Nanotubes, 
and Related Systems", ed. S. Saito, 
(AIP, New York, 2001), in press.\\
$[13]$ K. Harigaya and S. Abe, Phys. Rev. B {\bf 49}, 16746 (1994).\\
$[14]$ K. Harigaya, Phys. Rev. B {\bf 50}, 17606 (1994).\\
$[15]$ R. Saito, G. Dresselhaus, and M. S. Dresselhaus,
J. Appl. Phys. {\bf 73}, 494 (1993).\\
$[16]$ M. S. Dresselhaus and G. Dresselhaus,
Adv. Phys. {\bf 30}, 139 (1981).\\
$[17]$ K. Harigaya, A. Terai, Y. Wada, and K. Fesser,
Phys. Rev. B {\bf 43}, 4141 (1991).\\
$[18]$ K. Harigaya, J. Phys.: Condens. Matter
{\bf 3}, 4841 (1991).\\
$[19]$ K. Harigaya, Phys. Rev. B {\bf 53}, R4197 (1996).\\
$[20]$ T. Suzuki and K. Kaneko, Carbon {\bf 26}, 743 (1988).\\
$[21]$ M. Fujita, K. Wakabayashi, K. Nakada, and K. Kusakabe,
J. Phys. Soc. Jpn. {\bf 65}, 1920 (1996).\\
$[22]$ M. Fujita, M. Igami, and K. Nakada,
J. Phys. Soc. Jpn. {\bf 66}, 1864 (1997).\\
$[23]$ K. Nakada, M. Fujita, G. Dresselhaus, and M. S. Dresselhaus,
Phys. Rev. B {\bf 54}, 17954 (1996).\\

\pagebreak
\begin{flushleft}
{\bf Figure Captions}
\end{flushleft}

\mbox{}

\noindent
Fig. 1. Stacked nanographite with zigzag edges.
The bold and thin lines show the first and second
layers, respectively.  The stacking is the A-B type 
in (a), and it is the simple A-A type in (b).  
There are 24 carbon atoms in one layer.  The circles in (a)
show sites where the interlayer distance between 
carbon atoms is shortest.  The sites, A, B, ..., and 
F, are edge sites of the first layer, and the sites,
A', B', ..., and F', are edge sites of the second layer.
When there is a site potential at D (E or F) of
the first layer, another site potential is present
at D' (E' or F') of the second layer.

\mbox{}

\noindent
Fig. 2.  The absolute magnitude of the total magnetic
moment per layer of the model I as a function of $t_1$.
The stacking is the AB-type in (a), (b), and (c), 
and it is the AA-type in (d).  There is a site 
potential $E_s = -2 t$, (a) at the site D,
(b) at the site E, and (c) at the site F.  The site 
positions are displayed in Fig. 1 (a).  In (a), the 
onsite interaction is varied within $0.6t$ (closed squares) 
$\leq U \leq 1.8t$ (closed triangles).
The interval of $U$ between the series of the plots
is $\Delta U = 0.3t$.  In (b), it is varied within $1.0t$ 
(closed squares) $\leq U \leq 2.0t$ (closed triangles).
The interval of $U$ between the series of the plots
is $\Delta U = 0.25t$. In (c), it is varied within $0.6t$ 
(closed squares) $\leq U \leq 1.8t$ (closed triangles).
The interval of $U$ between the series of the plots
is $\Delta U = 0.3t$.   Finally, in (d), we display that 
there is no magnetization for the AA-stacking 
with a site potential at the D, E, or F.

\mbox{}

\noindent
Fig. 3.  The magnitude of the total magnetic moment
per layer of the model II as a function of $d$ and $U$.  
There is a site potential $E_s = -2 t$, (a) at the site D,
(b) at the site E, and (c) at the site F.  The site 
positions are displayed in Fig. 1 (a).  The interlayer
distance is $R=3.4$\AA.  The values of $U$ are $U=0.4t$ 
(filled squares), $0.8t$ (open squares), $1.2t$ (filled circles),
$1.6t$ (open circles), $2.0t$ (filled triangles),
respectively.

\mbox{}

\noindent
Fig. 4.  The magnitude of the total magnetic moment
per layer of the model II as a function of $d$ and $R$.
There is a site potential $E_s = -2 t$, (a) at the site D,
(b) at the site E, and (c) at the site F.  The site 
positions are displayed in Fig. 1 (a).  The strength
of the Coulomb interaction is $U=1.2t$.  The values 
of $R$ are $R=3.4$\AA (filled squares), 3.2\AA
(open squares), and 3.0\AA (filled circles), 
respectively.  All the plots within $0 \leq d \leq 0.5a$ 
overlap, so only the filled squares are seen.

\end{document}